 \documentclass[final,5p,times,twocolumn]{elsarticle}

\usepackage{amssymb}
\setcitestyle{square}
\setcitestyle{citesep={,}}
\biboptions{sort&compress}
\usepackage{amsmath}
\usepackage[hidelinks]{hyperref}
\usepackage{lipsum}
\usepackage{float}
\usepackage{graphicx}

\journal{Optics and Lasers in Engineering}

\begin{document}

\begin{frontmatter}



\title{Towards optimal multimode fiber imaging by leveraging input polarization and deep learning}
\author[a]{Jawaria Maqbool}
\affiliation[a]{organization={Department of Electrical Engineering, Syed Babar Ali School of Science and Engineering,  Lahore University of Management Science},
            addressline={Sector U, D.H.A Lahore, Cantt}, 
            city={},
            postcode={54792}, 
            state={Punjab},
            country={Pakistan}}

\author[b]{Syed Talal Hassan}
\affiliation[b]{organization={Department of Computer Science, Syed Babar Ali School of Science and Engineering,  Lahore University of Management Science},
            addressline={Sector U, D.H.A Lahore, Cantt}, 
            city={},
            postcode={54792}, 
            state={Punjab},
            country={Pakistan}}
\author[a,*]{M.Imran Cheema}    
\begin{abstract}
Deep learning techniques provide a plausible route towards achieving practical imaging through multimode fibers. The results produced by these methods are often influenced by physical factors like temperature, fiber length, external perturbations, and polarization state of the input light. Literature focuses on these different elements impacting deep-learning-enabled multimode imaging, yet the effects of input polarization remain under-explored. Here, we show experimentally that the state of polarization of light, being injected at multimode fiber input, affects the fidelity of reconstructed images from speckle patterns. Certain polarization states produce high-quality images at fiber output, while some yield degraded results. We have designed a conditional generative adversarial network~(CGAN) for image regeneration at various degrees of input light polarization. At a particular polarization state and with a thinner core multimode fiber, our network can reconstruct images with an average structural similarity index(SSIM) exceeding 0.9. Hence, in the case of multimode fibers that are held fixed, optimal imaging can be achieved by leveraging deep learning models with the input light polarization state, where the fidelity of images is maximum.  We also show that the model can be trained to image adequately for all input light polarization states when the fiber has bends or twists. We anticipate that our work will be a stepping stone toward developing high-resolution and less invasive multimode fiber endoscopes.
\end{abstract}


\end{frontmatter}



\section{Introduction}
\label{introduction}
Multimode fibers~(MMFs) can lead to practical endoscopes because they are thinner and less invasive than single-mode fiber bundles~\cite{hadley1965gastro,gu2014fibre,perperidis2020image}. The presence of numerous spatial modes in MMFs can be harnessed for image transmission. However, light waves propagating through different fiber modes interfere with each other to form speckles or random patterns at the fiber’s distal end. Hence, its properties resemble scattering or disordered media like fog, diffusers, and biological tissues that scramble the information to produce a speckle phenomenon. Extraction of data from speckle patterns is a challenging task. Primarily, three different strategies are generally employed for image reconstruction from speckles: optical phase conjugation~\cite{papadopoulos2012focusing,papadopoulos2013high}, computation of transmission matrix~\cite{popoff2010measuring,akbulut2013measurements,dremeau2015reference}, and deep learning~\cite{borhani2018learning}. Phase conjugation incorporates a complex interferometric method for measuring phase, and precise alignment is required between the camera and spatial light modulator. The transmission matrix~(TM), on the other hand, aptly describes the relationship between an MMF input and output. It helps to assimilate the information about light absorption, reflection, and transmission through the medium. The TM measurement requires both amplitude and phase information. The accurate phase computation needs a stable reference arm and nontrivial interferometric setup or a large number of input and output measurements~\cite{
metzler2017coherent,pellegrini2023focusing}. Moreover, phase values are very sensitive to external perturbations. Hence, one TM can only be used for the transmission state in which it is calculated~\cite{popoff2010measuring}. Recent research indicates that the challenges mentioned above can be addressed by applying deep learning techniques, leading to a more effective imaging process using MMFs ~\cite{rahmani2018multimode,fan2019deep}.

Previous deep learning works have shown various ways to improve MMF imaging in terms of accuracy, generalizability, and data requirements~\cite{zhu2021image,song2022deep,fan2019deep,resisi2021image,nguyen2021sensing,yu2021high,wang2022upconversion}. While extensive research has explored the effects of temperature, fiber length, and mechanical perturbations on deep learning based multimode fiber imaging, limited work is done in the past to investigate the impact of input polarization states in this domain. Prior work includes the characterization, statistics, and control of the polarization of light in multimode fibers. Due to random mode interference in multimode fibers, polarization mixing also occurs, which results in depolarized or partially polarized output~\cite{fridman2012modal}. On the other hand, it has been shown in~\cite{ploschner2015seeing} that the field distribution of some modes does not change during propagation through the fiber. Moreover, complete control of output polarization can be achieved using the eigenvectors and eigenvalues of the multimode fiber TM with orthogonal polarizations as a basis~\cite{xiong2018complete}. Lately, different polarization states at multimode fiber output have been exploited for image reconstruction and decomposition of vector modes ~\cite{song2022deep,xu2023multi}. Also, researchers have done channel classification and image regeneration for multiple wavelengths and input polarizations ~\cite{zhu2023deep}. They have used only linear polarization channels from 0 to 160~$^{\circ}$ with an interval of 20~$^{\circ}$. They have attained an average SSIM of 0.9413 for a multimode fiber with a large 200~$\mu$m core diameter and a high numerical aperture(NA) of 0.37. However, these works do not address the effect of diverse input polarization states on the fidelity of reconstructed images at the fiber output. Moreover, existing literature corroborates that multimode fibers with enlarged cores and greater acceptance angles enhance image quality under ideal conditions~\cite{zhong2023influences,papadopoulos2013high}.

Here, we devise an experimental and computational way to quantify input polarization impact on multimode fiber imaging. We acquire output data of speckles for nine different input polarization states, including horizontal, vertical,45~$^{\circ}$ linear, right and left circular, and four different elliptical polarizations at multiple MMF positions. We reconstruct original images from acquired speckle patterns using our designed conditional generative adversarial network~(CGAN). The model achieves stable training convergence within 1 hour. and enables accurate image reconstruction in just 9.4 ms. Our experimental results demonstrate that for specific input polarization states, the reconstructed images have higher fidelity based on measured quality metrics. In contrast, other launch polarizations produce lower-quality output images, establishing a distinct correlation between the input polarization and achievable spatial resolution. Furthermore, the highest SSIM that we achieve is 0.9046 at a specific polarization state for a fiber with 50~$\mu$m core diameter and NA of 0.22.  It is significant to mention here that we get a reasonably high value of SSIM even with a relatively thinner multimode fiber. 

Additionally, we find that our model trained for one polarization state at a particular fiber position gives poor reconstruction results for another polarization state. To reconstruct images at varying input polarizations and a specific fiber position, we merge an equal and small percentage of data from all nine polarization datasets to form one combined dataset. The model is trained on this dataset and tested on unseen data of each polarization state. This procedure is carried out separately for two fiber positions. Furthermore, we integrate subsets of eighteen datasets for both fiber positions. After training on this super set, our CGAN model can accurately reconstruct images for unseen data of all polarization states of two fiber positions under consideration. Hence, our work highlights that the input light polarization state affects the accuracy of reconstructed images from speckles at the multimode fiber output, and it can be harnessed in two ways: 1) For a fixed MMF orientation, input polarization can be set to a degree where we get optimal imaging results,2) But in scenarios, where fiber position can change, we must train our model on the data measured while constantly changing fiber position and input light polarization state. In this way, we can get satisfactory reconstruction results for any input polarization state.

We now describe the rest of the paper. Section~\ref{sec:Exp_setup} details the experimental setup for data collection, followed by Section~\ref{sec:Data_acq}, in which we describe the data acquisition procedure. Section~\ref{se:DL_framework} is dedicated to our deep learning framework, where we introduce its architecture, training processes, and integration with the data gathered in the previous sections. Section~\ref{sec:Acess_SOP} presents our methodology for evaluating the input polarization impact on the reconstructed images' quality and offers insights into the system’s sensitivity to polarization variations. Section~\ref{sec:model_general} explains our model's ability to reconstruct for diverse input polarizations. Finally, Section~\ref{sec:Concl} summarizes our findings and highlights potential avenues for future research.

\section{Experimental setup} \label{sec:Exp_setup}
\begin{figure*}
\centering\includegraphics[width=13.90cm]{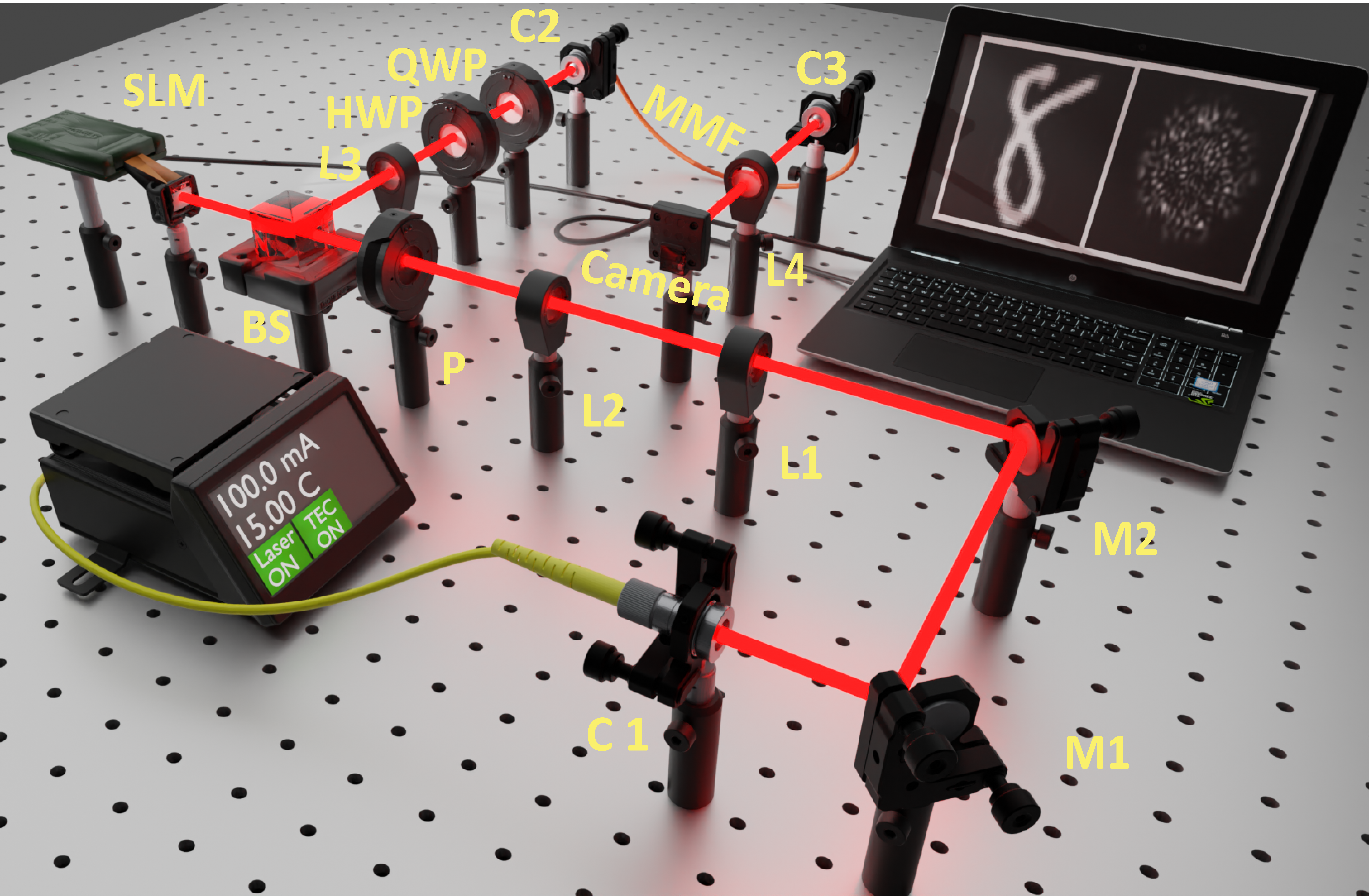}
\caption{Experimental schematic illustrating our multimode fiber~(MMF) imaging process using a combination of input polarization and deep learning techniques. The illustration depicts only one example of MMF positioning. For other fiber positions, curling or bending the fiber is done arbitrarily. C:Collimator, M:Mirror, L:Lens, P:Linear polarizer, BS:Beam splitter, SLM:Spatial light modulator, HWP:Half wave plate, QWP:Quarter wave plate, and MMF:Multimode fiber.}
\label{setup}
\end{figure*}

The experimental setup schematic is illustrated in Fig.~\ref{setup}. We utilize a 633~nm continuous-wave laser diode~(Eagleyard GC-02940) operated via  Thorlabs CLD1015 controller. After reflection by mirrors, the laser light is collimated through a telescopic system comprising two lenses with focal lengths of 500~mm and 100~mm. A polarizer is placed after the telescopic system to achieve horizontal polarization for optimal phase modulation with the HOLOEYE Pluto 2.0 spatial light modulator~(SLM). The polarized laser beam is then directed onto a 50/50 beam splitter~(BS). Half of the beam is transmitted towards the SLM, while a beam blocker blocks the remaining half. Once reflected by the SLM, the phase-modulated light passes through the BS and is imaged by lens 3 onto collimator 2, which in turn focuses the image of the phase-modulated light onto the input of a multimode fiber. The multimode fiber has core and cladding diameters of 50~$\mu$m and 125~$\mu$m, respectively, with a length of 1~m and a numerical aperture (NA) of 0.22. Before the fiber input, a half-wave plate~(HWP) and a quarter-wave plate~(QWP) are positioned to attain any desired state of polarization~(SOP). The multimode fiber converts all the information the laser light carries into a speckle pattern. The speckle pattern emerging from collimator 3 is imaged by lens 3 onto a Thorlabs DCC1545M CMOS camera with a resolution of 1280$\times$1024 pixels.
\begin{figure*}
\centering\includegraphics[width=13.90cm]{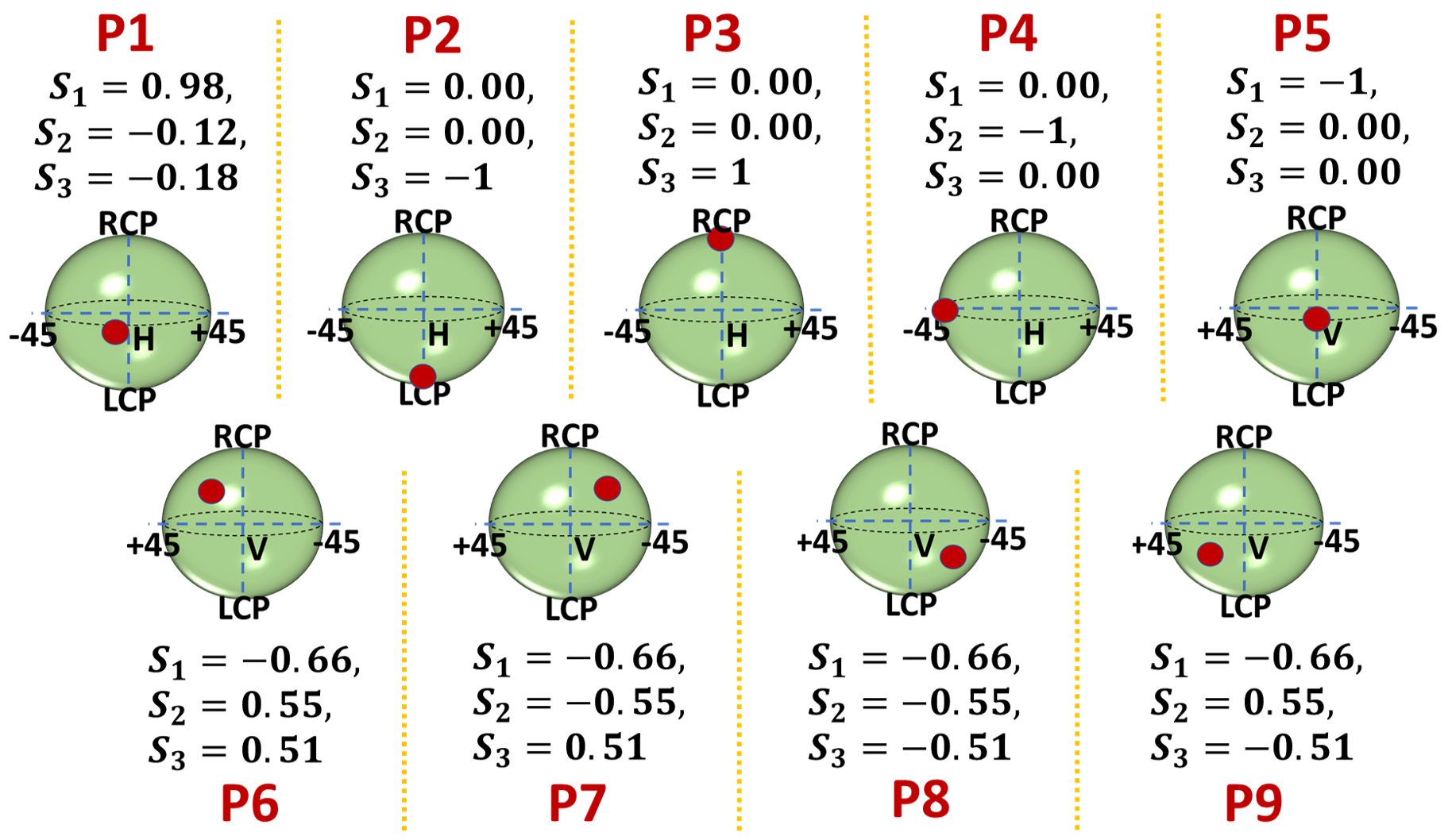}
\caption{The nine polarization states used in this work are depicted on the Poincaré sphere along with the three Stokes parameters (S$_1$,S$_2$,S$_3$)~\cite{kliger_2012} that are measured experimentally. }
\label{poincere}
\end{figure*}
\section{Data acquisition for different polarization states} \label{sec:Data_acq}
Initially, we place the multimode fiber in a particular position 1 and its orientation remains fixed for all measurement sets. Fixing the position is essential as speckle patterns change with a change in the orientation of the fiber~\cite{resisi2021image}. A state of polarization~(SOP) at the fiber input is set with HWP and QWP while observing SOP on Thorlabs’s PAX1000IR1/M polarimeter. We choose nine different SOPs comprising linear, circular, and elliptical polarizations, as shown in Fig.~\ref{poincere}. The Stoke’s parameters indicated in Fig.~\ref{poincere} are measured using the polarimeter.  For each input SOP, a computer sends the Modified National Institute of Standards and Technology (MNIST) data of 50,000 handwritten digits on SLM. The images are of size 28$\times$28 pixels and are up-sampled to 64$\times$64 pixels before being sent on SLM. The light reflected from SLM now contains images of handwritten digits. After passing through MMF, this light produces speckle patterns that are recorded by the computer connected to the camera. The speckle patterns saved on the system are cropped to dimensions of $256\times256$ pixels. The process of speckle data collection for 50,000 images takes approximately 
\begin{figure*}[hbt!]
\centering\includegraphics[width=14cm]{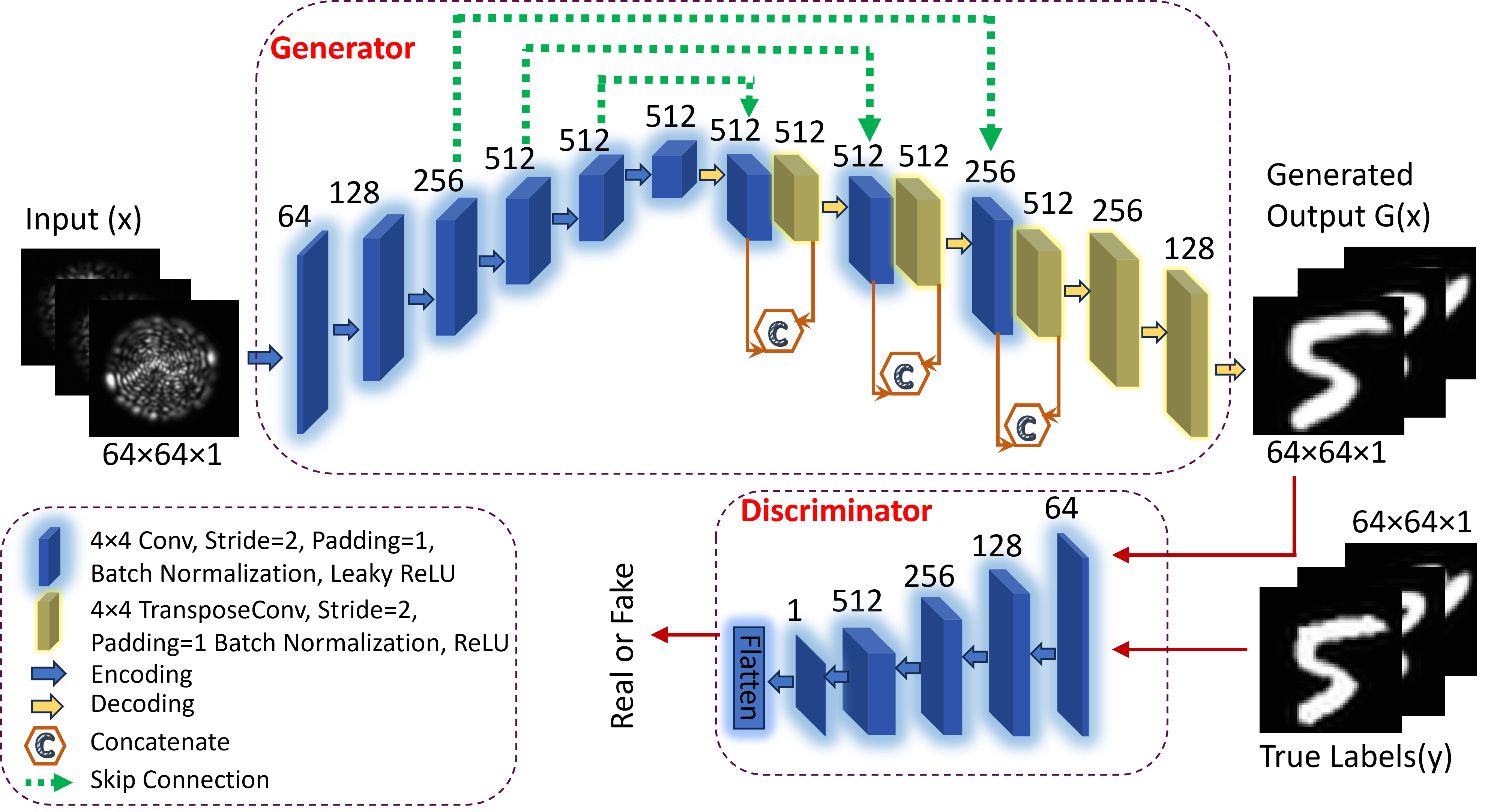}
\caption{The structure of our designed conditional generative adversarial network~(CGAN).}
\label{cgan}
\end{figure*}
23 hours. We perform this procedure for nine input polarization states, resulting in the formation of nine different datasets. To further gauge the input polarization effect on imaging accuracy through MMF, we change the fiber position and repeat the methodology of acquiring nine datasets for nine different polarization states.
\section{Deep learning framework for image reconstruction} \label{se:DL_framework}
After the formation of data sets at various distinct input polarization states, the next step is reconstructing original images from speckle patterns. For this, we design a conditional generative adversarial network, as shown in detail in Fig.~\ref{cgan}. The generator is a U-Net-type architecture with an encoder, decoder, and skip connections. We first down-sample the speckle patterns to size 64$\times$64$\times$1 and apply them as the input to the generator. The generator is enabled with robust feature extraction capabilities due to several convolution and deconvolution layers. In addition, the skip connections allow weight sharing and preserve feature information across different network layers. The output has the same resolution as the input. The discriminator is composed of five convolution layers and one flattened layer. The generator’s output, concatenated with the true label, is employed as input to the discriminator, which works on classifying a patch in an image as real or fake.

CGAN has been used previously for reconstructing images from speckle patterns produced by multimode fibers ~\cite{yu2021high,wang2022upconversion}. The highest average SSIM reported previously is 0.8686~\cite{song2022deep}. In contrast to previous works, we use binary cross entropy~(BCE) loss for the discriminator and an amalgam of mean squared error~(MSE) and mean absolute error~(MAE) for the generator. The MSE loss function minimizes the difference between real and generated data. It also overcomes the problem of vanishing gradients, resulting in stable training and high-fidelity results~\cite{mao2017least}. MAE loss aids in the regeneration of low-frequency details. The hybrid loss function for a CGAN is the weighted sum of generative and discriminative loss~(Eq.\eqref{Eq:1}). We define the discriminator and generator loss for our designed model in Eq.\eqref{Eq:2} and Eq.\eqref{Eq:3}, respectively:
\begin{equation}
\label{Eq:1}
\mathcal{L}_{CGAN} =\mathcal{L}_{Gen} +\mathcal{L}_{Disc},
\end{equation}
\begin{equation}
\label{Eq:2}
\mathcal{L}_{Disc} =\lambda_1[l_1(D(y,x),1]+\lambda_1[l_1(D(G(x),x),0],
\end{equation}
\begin{equation}
\label{Eq:3}
\mathcal{L}_{Gen} =l_2[D(G(x),x),1]+\lambda_2[l_3((G(x),y)],
\end{equation}
where $G()$ and $D()$ are the generator and discriminator functions, respectively. The speckle pattern inputs are represented by $x$, while the true labels are denoted by $y$. We use $l_1$ for BCE, $l_2$ for MSE, and $l_3$ for MAE. To optimize the model's performance, we incorporate weighting factors, $\lambda_1$ and $\lambda_2$, set at values of 100 and 0.5, respectively, to effectively balance the MAE and BCE losses. Out of 50000 pairs of speckle-MNIST digits in all polarization datasets, we reserve 5000 pairs for testing. For the remaining 45000 pairs, 85$\%$ are kept for training, and the rest 15$\%$ are used for validation. For each data set, the CGAN model takes 1 hour to train for 80 epochs, and the training process remains stable with no mode collapse. The inference time for each reconstructed image is 9.4~ms. We realize the data collection using a Python 3.10.12 environment. Furthermore, we utilize the PyTorch framework for building, training, and testing the model. The whole mechanism of deep learning is accelerated by the NVIDIA Tesla V100 Tensor Core GPU.

\begin{figure}
	\centering 
	\includegraphics[width=0.48\textwidth, angle=0]{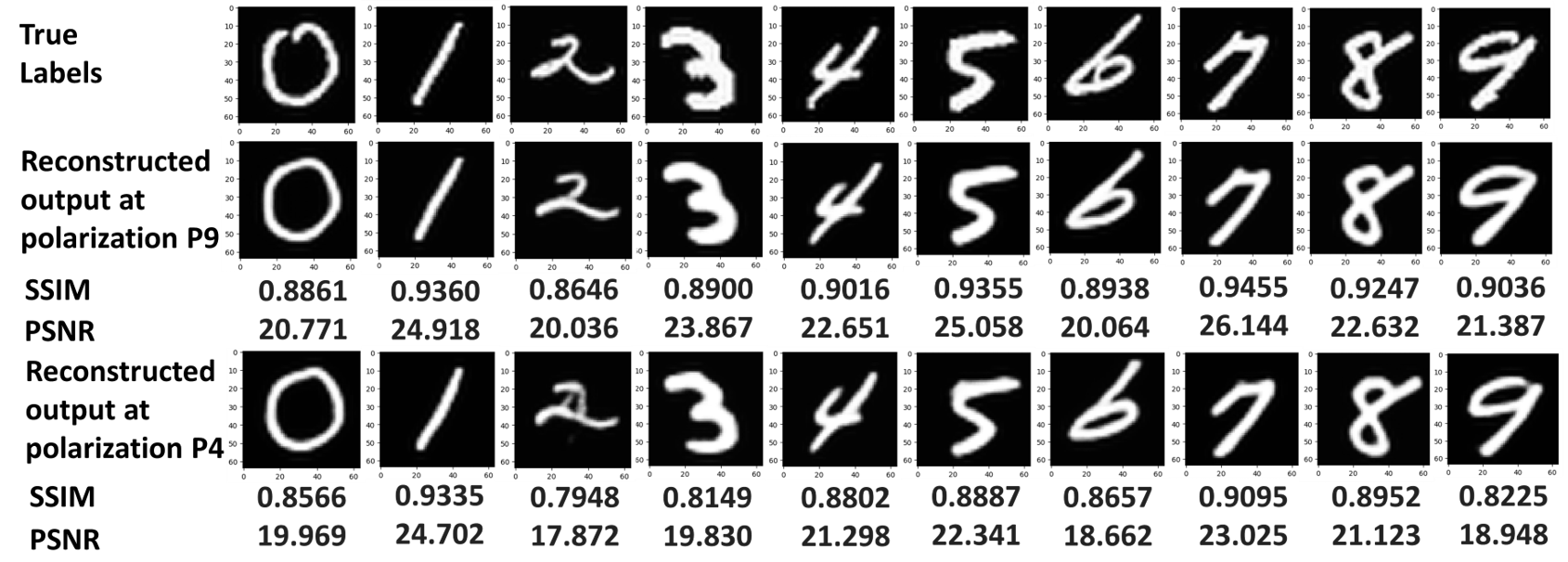}	
	\caption{Representative image reconstruction results at two different polarization states for position 1 of MMF. At P9~(elliptical), the average SSIM and PSNR are maximum, while these metrics have the lowest average values at P4~(45-degree). Please see Fig.~\ref{poincere} P9 and P4 Stokes parameters.}
\label{pos1}
\end{figure}

\begin{figure}
	\centering 
	\includegraphics[width=0.48\textwidth, angle=0]{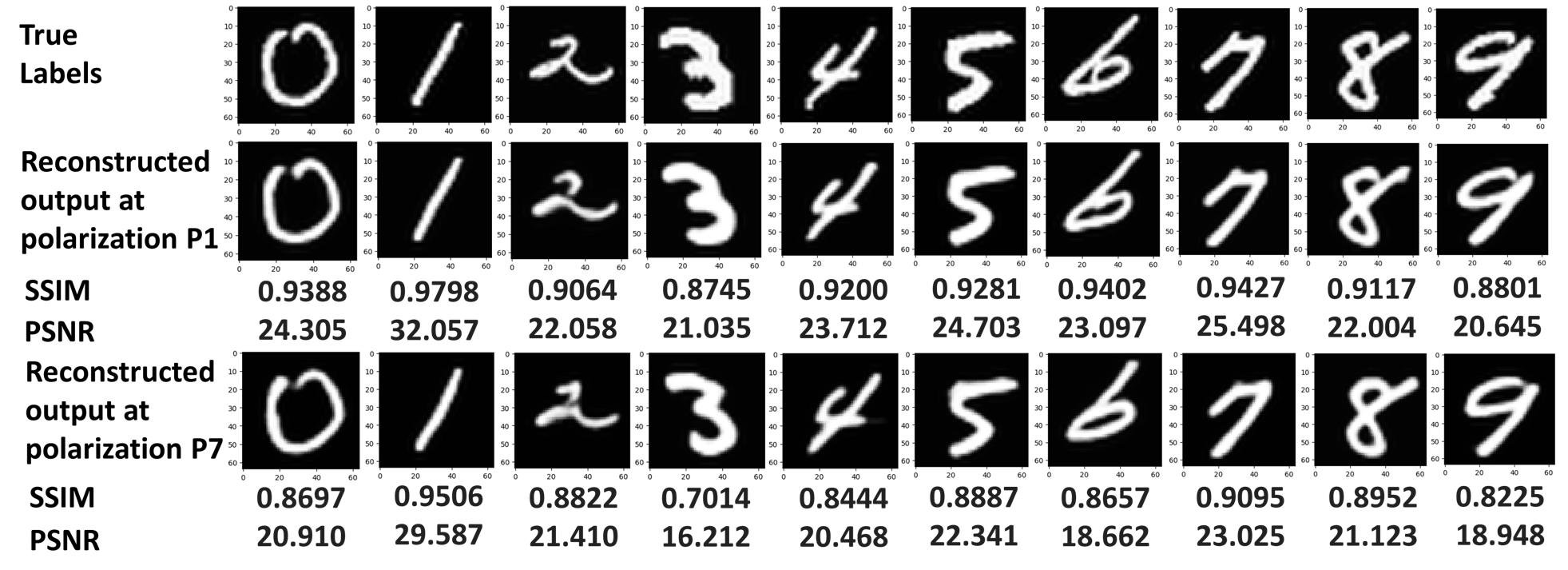}	
	\caption{Representative reconstruction results at two different polarization states for position 2 of MMF. At P1~(nearly horizontal), the average SSIM and PSNR are maximum, while these metrics have the lowest average values at P7~(elliptical). Please see Fig.~\ref{poincere} for P1 and P7 Stokes parameters.}
\label{pos2}
\end{figure}

\begin{figure}
	\centering 
	\includegraphics[width=0.48\textwidth, angle=0]{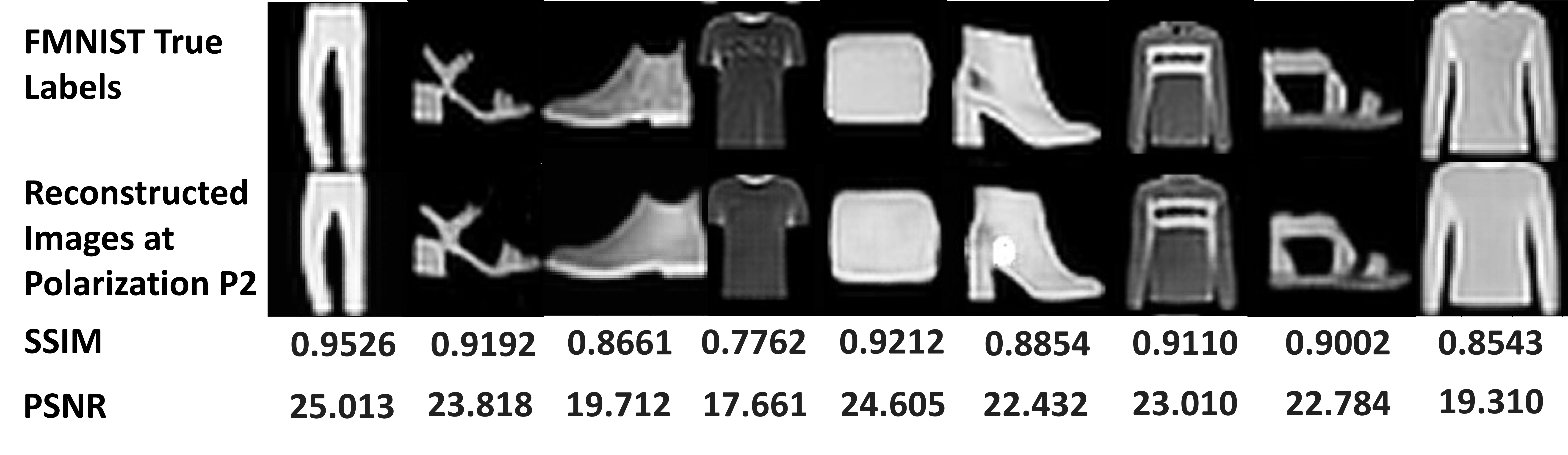}	
	\caption{Representative image reconstruction results for Fashion MNIST data and left circularly polarized light(P2) at multimode fiber input.}
\label{fm}
\end{figure}

We train and evaluate our CGAN model for all compiled polarization data sets at fiber positions 1 and 2. We use SSIM and peak signal-to-noise ratio~(PSNR) as evaluation metrics for our restored images. SSIM compares the similarity between reconstructed digits and true labels based on their luminance, contrast, and structure. Its value varies between 0 and 1. An SSIM value around zero means no similarity between the two images, while a value closer to 1 denotes that the images are almost identical. Its expression is given by:
\begin{equation}
SSIM =
 \frac{(2\mu_x\mu_y +C_1)(2\sigma_{xy}+C_2)}{(\mu_x^2+\mu_y^2+C_1)(\sigma_x^2+\sigma_y^2+C_2)},
\end{equation}
where $\mu_x$ and $\mu_y$ refer to the mean value over a window in images $x$ and $y$, respectively. $\sigma_x$ and $\sigma_y$ are standard deviations over a window of $x$ and $y$. $\sigma_{xy}$ is the covariance across a window between image $x$ and image $y$ while $C_1$ and $C_2$ are constants.
Another metric that we have used is PSNR, which is the ratio between the maximum pixel value of the ground truth image~($I_{max}$) and the root mean squared error~(RMSE) and is given by:
\begin{equation}
PSNR =
 20log_{10}\dfrac{I_{max}}{RMSE}.
\end{equation}
RMSE is determined between the pixel values of the original and the predicted images. The higher the value of PSNR, the better the quality of reconstructed images. Some of the reconstruction results from our designed CGAN are given in Fig.~\ref{pos1} and Fig.~\ref{pos2} for fiber positions 1 and 2, respectively. For brevity, the regenerated images for each fiber position are displayed for only two polarization states: one where SSIM and PSNR attain their respective maximum values and the other where they reach their minimum values. This choice is made to show a proper difference in image fidelity at these polarization states. As can be observed in Fig.~\ref{pos1}, at polarization P9~(elliptical), the images are closer to true labels~(have high SSIM and PSNR) as compared to polarization P4~(45-degree). P4 has relatively poor image regeneration results, especially for digits 2 and 3. The same can be apprehended from the results of Fig.~\ref{pos2}, where average PSNR and SSIM attain their highest values at P1~(nearly horizontal) and lowest at P7~(elliptical). It is important to highlight that in our experiments, we deploy the MNIST dataset only because our purpose is to specifically illustrate the impact of input polarization states on reconstruction fidelity. Our model also demonstrates good performance for Fashion MNIST data reconstruction. The average SSIM, in this case, surpasses the value of 0.88 at a specific polarization state. The results for Fashion MNIST data are shown in Fig.~\ref{fm}.
\section{Assessing the input polarization effect}\label{sec:Acess_SOP}
\begin{figure}[H]
	\centering 
	\includegraphics[width=0.48\textwidth, angle=0]{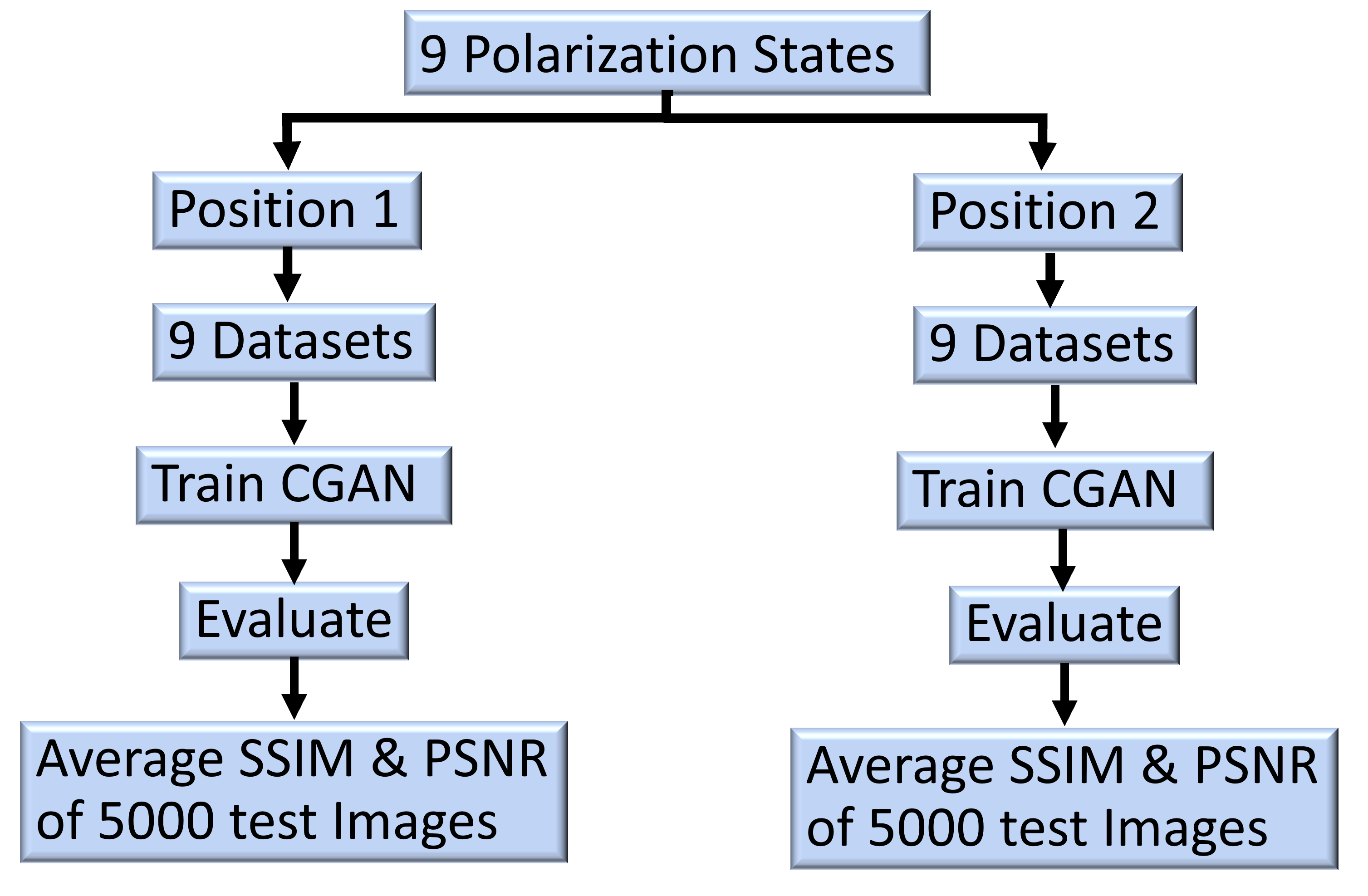}	
	\caption{A flow chart showing the steps we follow to access input polarization effect}
\label{flow}
\end{figure}
We record eighteen different data sets for two distinct positions of the multimode fiber at nine varying input polarization states shown in Fig.~\ref{poincere}. We then train our designed CGAN for these data sets, followed by the model evaluation for 5000 unseen test images. An overview of this process is given in  Fig.~\ref{flow}.

\begin{figure}
	\centering 
	\includegraphics[width=0.48\textwidth, angle=0]{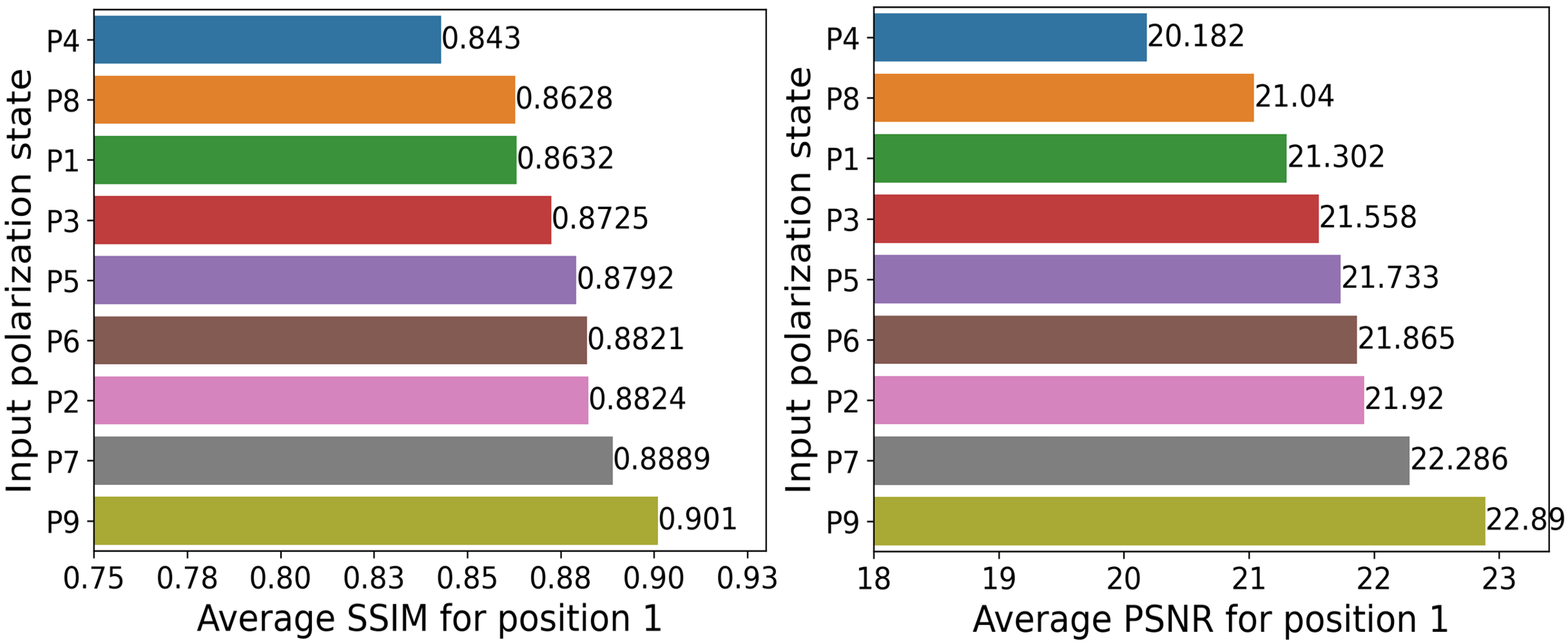}	
	\caption{The average SSIM and PSNR variations of 5000 unseen test images for individual polarization states when the fiber is fixed in position 1.}
\label{plot_pos1}
\end{figure}

\begin{figure}
	\centering 
	\includegraphics[width=0.48\textwidth, angle=0]{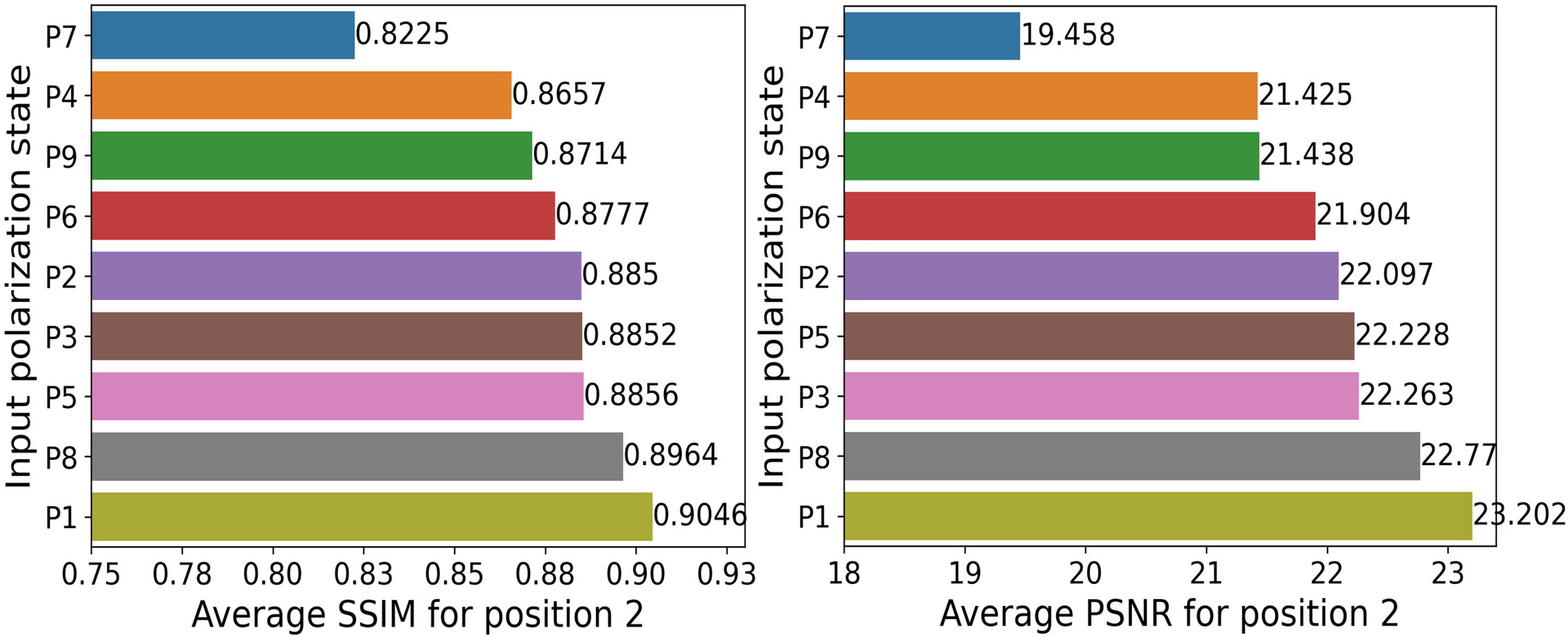}	
	\caption{The average SSIM and PSNR variations of 5000 unseen test images for individual polarization states when the fiber is fixed in position 2.}
\label{plot_pos2}
\end{figure}

The obtained average SSIM and PSNR for every data set are shown in Figs.~\ref{plot_pos1} and \ref{plot_pos2}. The varying magnitudes of these bar graphs illustrate that SSIM and PSNR change with input polarization states. For position 1, the maximum average SSIM of 0.9010 and PSNR of 22.89 are attained at elliptical polarization. At the same time, the minimum SSIM of 0.8430 and PSNR of 20.182 are obtained for linearly polarized light at 45 degrees. The percentage difference between the smallest and largest SSIM is 6.65$\%$, while for PSNR, this variation is 12.5$\%$. For position 2, the highest average SSIM of 0.9046 and PSNR of 23.202 are achieved for nearly horizontally polarized light. The lowest SSIM of 0.8225 and PSNR of 19.458 are obtained when the input light is vertically polarized. In this case, the percentage difference between the two extremities is 9.5$\%$ for SSIM and 17.55$\%$ for PSNR. We find that the deviation between evaluation metrics' values for some polarization states is more significant than others.

It can also be inferred from the plots that the effect of different polarization states changes with a change in the fiber position. For example, at P1 and position 2, SSIM of 0.9046 and PSNR of 23.202 are high, but for position 1 and the same polarization state, these metrics have reduced to 0.8632 and 21.302, respectively. This is because modal distribution changes with the bending or twisting of the fiber.

We repeat the data collection, model training, and testing procedure twice at each of the nine polarization states and for individual fiber positions. This is done to ensure the capture of the persistent impact of different polarization states on the fidelity of reconstructed images. When evaluated across various input polarization states, we observe that the percentage difference in SSIM and PSNR values for reconstructed images exhibit consistent results with only 1-2$\%$ marginal fluctuations. This means that if SSIM and PSNR are minimum at P4~(45-degree linear) compared to other polarization states, it will always be lowest, no matter how many times we repeat this process while keeping the position fixed.

Physically, the relationship of reconstruction of images with input polarization states can be elaborated in the following way. When light with a specific polarization state is launched to any fiber mode, it spreads to other modes. Due to modal coupling, polarization scrambling also occurs, resulting in different polarization states at the inputs and outputs of all modes. Moreover, higher-order modes suffer from higher attenuation than lower-order modes. For an arbitrary polarized~($p$) input $|\phi\bigr\rangle$ the output field is $|\psi\bigr\rangle = t_{p}|\phi\bigr\rangle$, where $t_{p}$ is the transmission matrix for $p$ polarized input. The total intensity of this polarization state is $\bigr\langle\psi|\psi\bigr\rangle =\bigr\langle\phi|t_{p}^\dag t_{p}|\phi\bigr\rangle $. The transmission range achieved in this state is defined by the eigenvalues of $t_{p}^\dag t_{p}$. The maximum energy that can be maintained in the same state of polarization is given by the largest eigenvalue, while the maximum energy retained in orthogonal SOP is defined by the smallest eigenvalue~\cite{xiong2018complete}. The larger eigenvalues and their associated eigenvectors get their contribution from lower-order modes, leading to maximum transmission. The eigenvectors corresponding to smaller eigenvalues are influenced by higher-order modes, resulting in reduced transmission. Also, input wavefronts change due to data sent on SLM. For some states of input polarization, the eigenvectors of most of the wavefronts from SLM coincide with greater eigenvalues of $t_{p}^\dag t_{p}$, contributing to the maximum transmission of these wavefronts. This eventually improves the fidelity of reconstructed images as most of the input information is retained while propagating through fiber. On the contrary, for certain input SOPs, eigenvectors of input wavefronts correspond to smaller eigenvalues, causing the attenuation of input information. SSIM and PSNR will be low in these cases.  Also, due to variations in mode and polarization coupling along with changes in the transmission matrix and its eigenvalues for various fiber positions, the influence of input polarization differs between the two fiber positions.
\section{Reconstruction for varying input polarization states}\label{sec:model_general}
The input polarization effect can be harnessed in two ways: (a) The input polarization is set at a degree where we get the optimal imaging results for endoscopic applications where multimode fiber length is small and is not bent or twisted while imaging~\cite{turtaev2018high}. (b) In the case of long-length endoscopes inserted deeply in the body, a dynamically perturbed multimode fiber should also be trained or calibrated for a diverse range of input polarization degrees. This approach ensures consistently satisfactory imaging results regardless of the input polarization degree.
\begin{figure}
	\centering 
	\includegraphics[width=0.48\textwidth, angle=0]{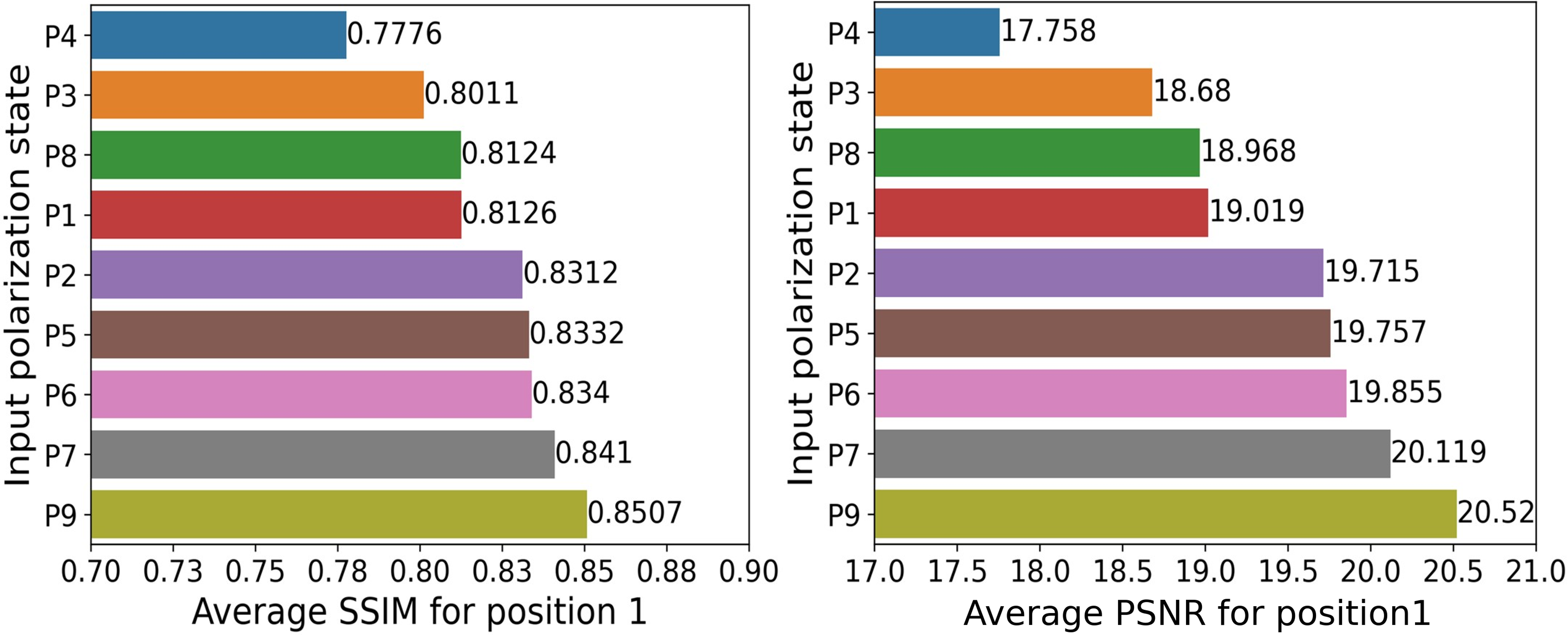}	
	\caption{The average SSIM and PSNR variation of 5000 unseen test images for individual polarization states when the fiber is fixed in position 1. First, the model is trained on combined data that contains $10\%$ images from every polarization state dataset and then is tested for all polarization states. }
\label{pos1_combined}
\end{figure}
\begin{figure}
	\centering 
	\includegraphics[width=0.48\textwidth, angle=0]{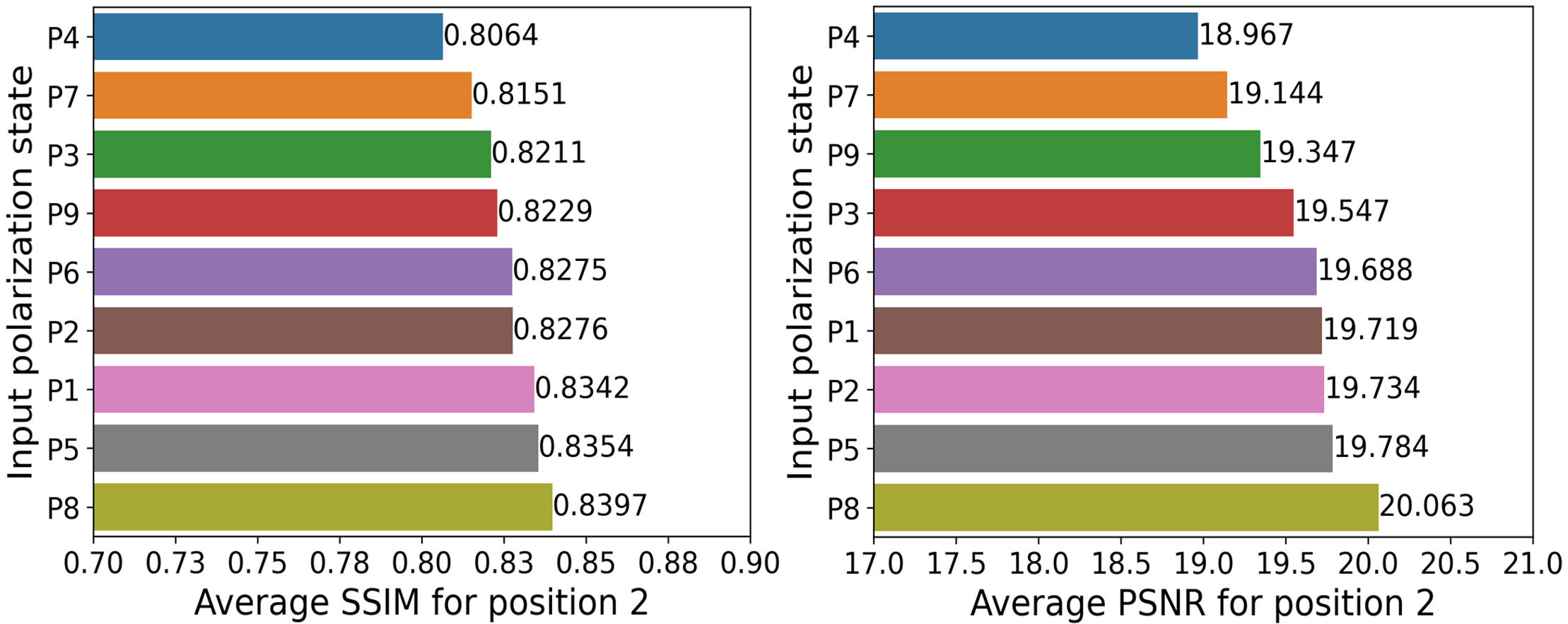}	
	\caption{The average SSIM and PSNR variation of 5000 unseen test images for individual polarization states when the fiber is fixed in position 2. First, the model is trained on combined data that contains $10\%$ images from every polarization state dataset and then is tested for all polarization states. }
\label{pos2_combined}
\end{figure}

Initially, we use the weights of one input polarization state to reconstruct the test images of another input polarization data. Not surprisingly, we get poor results as the average SSIM and PSNR remain below 0.2 and 8, respectively. We then combine 10$\%$ training data of each of the nine input polarization states to form two separate datasets for fiber positions 1 and 2. We train our designed CGAN model on these collective sets of images and test for 5000 unseen images of each polarization state and fiber position. The average SSIM and PSNR of nine polarization states tested on weights of combined datasets are plotted in Fig.~\ref{pos1_combined} and Fig.~\ref{pos2_combined} for positions 1 and 2, respectively. The SSIM and PSNR values are reasonable in this case. SSIM is not less than 0.77, and PSNR is greater than 17 for position 1. Likewise, for position 2 combined data, SSIM and PSNR remain above 0.8 and 18, respectively.
\begin{figure}
	\centering 
	\includegraphics[width=0.48\textwidth, angle=0]{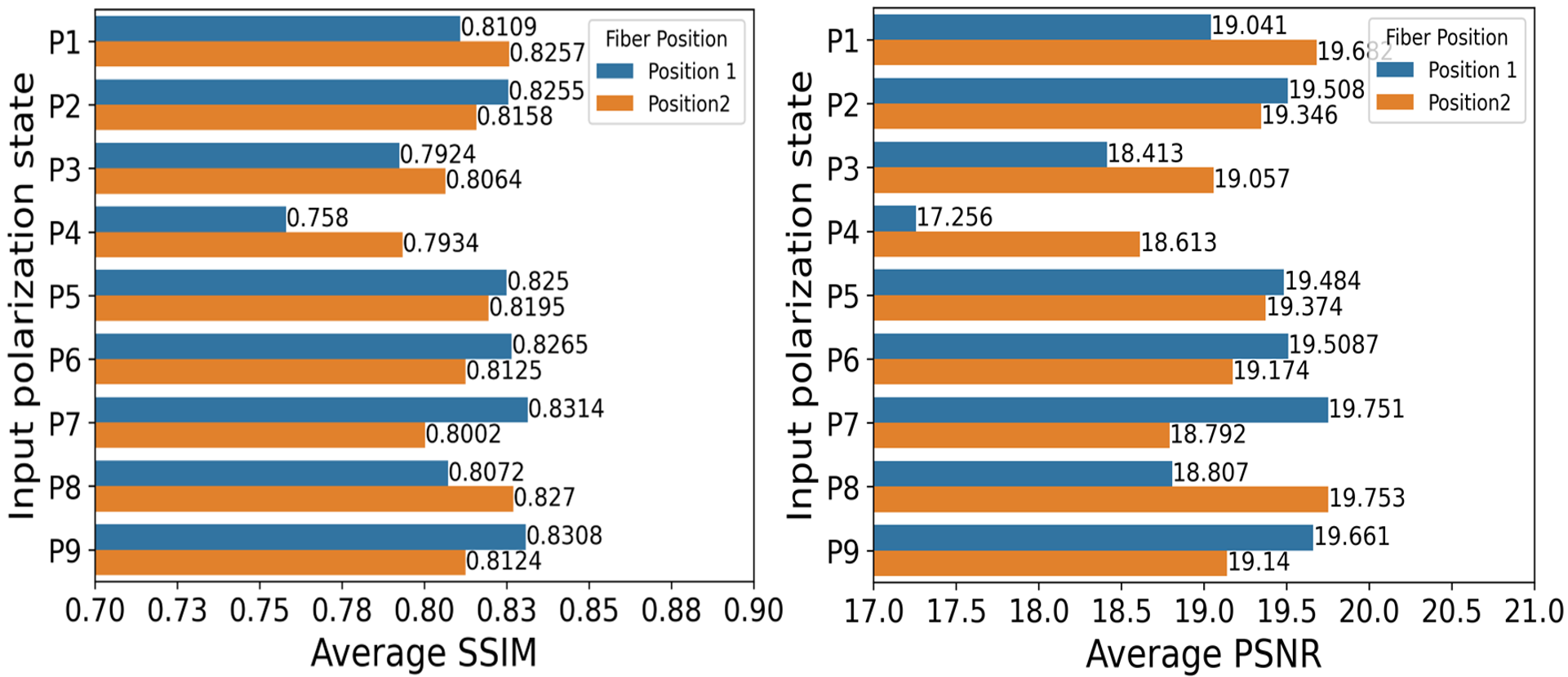}	
	\caption{The average SSIM and PSNR variations of 5000 unseen test images for individual polarization states for positions 1 and 2. First, the model is trained on combined data that contains 5$\%$ images from each polarization state dataset of both positions and then is tested for all polarization states.}
\label{pos1_pos2}
\end{figure}
As a final step where an MMF can image well for any input polarization state~(P1-P9) and position~(1 or 2), we pick up 5$\%$ training data from 18 datasets and integrate them to form one super set. After training on this set, we test our designed CGAN model for unseen images of all polarization states at both fiber positions. In this case, the average SSIM and PSNR are shown in Fig.~\ref{pos1_pos2}. It is noticeable from this bar graph that SSIM does not degrade more than 0.75 while PSNR is greater than 17. These reasonable metric values signify successful image reconstruction for any input polarization state. This process indicates that the CGAN model should be trained on a dataset obtained while dynamically changing the input polarization state and fiber positions for imaging at varying input polarizations and fiber positions. Table~\ref{Tab:1} encapsulates the mean and standard deviation of evaluation metrics when training is done separately on each polarization dataset and for a combination of different polarization and position datasets. As can be seen, mean values are higher after training on individual polarization datasets. However, as discussed previously, the model trained for one polarization does not reconstruct well for another polarization state. The mean metric values are relatively low for a combination of datasets. The small standard deviation values in case of varying input polarizations depict that the model can image well for any polarization state.

\section{Conclusion}\label{sec:Concl}
We demonstrate experimentally the influence of diverse input light polarization states on the accuracy of image reconstructions from speckle patterns at the multimode fiber output. Specifically, we have established a clear correlation between the input polarization states and the variation in the average structural similarity index (SSIM) and peak signal-to-noise ratio (PSNR) across a set of 5000 unseen images. These metric values are high for some input polarization states and relatively lower for others.  Furthermore, we have exhibited this polarization impact for two distinct multimode fiber positions. Significantly one of the polarization states gives a reasonably high SSIM exceeding 0.9 at multimode fiber output with a thin 50~$\mu$m core diameter. Each image is reconstructed in just 9.4~ms due to our designed CGAN  model that trains in only 1~hour, leading to real-time imaging through MMFs. Hence, for optimal imaging through a multimode fiber with an invariant position, we should incorporate that polarization state where SSIM is maximum. Moreover, for the cases where input polarization and fiber position are changing, we train the model for combined data from all polarization states and fiber positions. Through this approach, we demonstrate the feasibility of satisfactory imaging through multimode fibers for various polarization states and fiber positions. Our work can be extended to explore the influence of input polarization on multimode fiber-optic communication systems. Furthermore, this research is transferable to imaging applications through challenging mediums such as fog and biological tissues. We believe that our work can significantly contribute to developing compact and high-resolution endoscopes that do not require traditional lenses.
\renewcommand{\arraystretch}{0.9}
\begin{table*}[hbt!]
\centering
\begin{tabular}{l c c c c}
\hline
\rule{0pt}{10pt}
\textbf{Fiber Orientation} & \textbf{Mean SSIM} & \textbf{SSIM standard} & \textbf{Mean PSNR} & \textbf{PSNR standard} \\
\textbf{} & \textbf{} & \textbf{deviation} & \textbf{} & \textbf{deviation}\\
\hline\\
\textbf{Position 1, separate} & 0.8750 & 0.024 & 21.640 & 0.9684\\
\textbf{polarization states}                                    \\
\rule{0pt}{3pt}\\
\hline \\
\textbf{Position 2, separate} & 0.8771 & 0.023 & 21.865 & 1.067\\
\textbf{polarization states}                                  \\
\rule{0pt}{3pt}\\
\hline\\
\textbf{Position 1, varying} & 0.8215 & 0.0169 & 19.376 & 0.8475\\
\textbf{polarization states}                                   \\
\rule{0pt}{3pt}\\
\hline\\
\textbf{Position 2, varying} & 0.8255 & 0.0105 & 19.554 & 0.3437\\
\textbf{polarization states}                                    \\
\rule{0pt}{3pt}\\
\hline\\
\textbf{Position 1 $\&$2 varying} & 0.8122 & 0.0171 & 19.134 & 0.6138\\
\textbf{polarization states}                                   \\
\rule{0pt}{3pt}\\
\hline\\
\end{tabular}
\caption{Table showing mean metric values and standard deviation across various scenarios. Rows 1 and 2 display metrics for separate training and testing on individual polarizations and fiber positions. Rows 3 and 4 present results when combining 9 polarization datasets for each fiber position during training and testing across different polarizations of the respective positions. The last row illustrates training on a combination of 18 datasets from both fiber positions, with subsequent testing on various polarization states.}
\label{Tab:1}
\end{table*}
\bibliographystyle{elsarticle-num}
\bibliography{paper5}




\end{document}